\documentclass{article}
\pdfoutput=1
\usepackage{arxiv}
\usepackage[utf8]{inputenc}
\usepackage[english]{babel}

\usepackage{lmodern}
\usepackage{amssymb}
\usepackage{mathtools}
\usepackage{color}
\usepackage{float}

\usepackage[pdftex]{graphicx}
\usepackage[caption=false,font=footnotesize]{subfig}
\usepackage{array}
\usepackage{url}
\usepackage{booktabs} 
\usepackage{multirow}
\usepackage{colortbl}
	\definecolor{lgray}{gray}{0.75}
	\definecolor{dgray}{gray}{0.55}
	
	\usepackage{enumitem}

	\usepackage{color}	

\usepackage{algorithmicx}
\usepackage[noend]{algpseudocode}
		\algrenewcommand\algorithmicindent{2.0em}%
	\algnewcommand\Let[2]{\State #1 $\gets$ #2}
	\algnewcommand\AND{\ \textbf{and}\ }
	\algnewcommand\OR{\ \textbf{or} \ }
	\algnewcommand\algorithmicinput{\textbf{Input:}}
	\algnewcommand\Input{\item[\algorithmicinput]}
	\algnewcommand\algorithmiccompute{\textbf{Compute:}}
	\algnewcommand\Compute{\item[\algorithmiccompute]}

	\DeclareGraphicsExtensions{.jpg, .pdf, .png}
\graphicspath{	{./Figures/}
				}			    

\usepackage{hyperref}

\usepackage[backend=biber, style=ieee, doi=false, isbn=false, url=false]{biblatex} 
	\newcommand{\secref}[1]{\hyperref[#1]{Section~\ref{#1}}}
	\newcommand{\figref}[1]{\hyperref[#1]{Figure~\ref{#1}}}
	\newcommand{\tblref}[1]{\hyperref[#1]{Table~\ref{#1}}}
	\newcommand{\algoref}[1]{\hyperref[#1]{Algorithm~\ref{#1}}}

	\newcommand{\D} {\mathcal{D}}
	\newcommand{\G} {\mathcal{G}}
	\newcommand{\R} {\mathbb{R}}
	
	\newcommand{\M} {\mathcal{M}}

	\newcommand{\udvar}[1]{\underline{{#1}}}
	\newcommand{\ddvar}[1]{\underline{\underline{{#1}}}}
	\newcommand{\tdvar}[1]{\underline{\underline{\underline{{#1}}}}}

	\newcommand{\q}{\udvar{q}}
	\newcommand{\Q}{\tdvar{Q}}
	\newcommand{\T}{\ddvar{T}}
	\newcommand{\B}{\tdvar{B}}

	\newcommand{\Argmin}[1]{\ensuremath{\mathrm{Arg}\underset{#1}{\mathrm{min}\,}}}

	\def\eg{e.g.,\ }
	\def\cf{\textit{cf.\ }}
	\def\ie{\textit{i.e.,\ }}
	\def\am{a.m.}

\renewcommand{\today}{February 2017}
	
\begin{document}

\title{Combining traffic counts and Bluetooth data for link-origin-destination matrix estimation in large urban networks: The Brisbane case study}
\author{Gabriel Michau\\
\textit{ETH Zurich }\\
Zurich, Switzerland \\
\texttt{gmichau@ethz.ch}
\And
Nelly~Pustelnik\\
Univ Lyon, Ens de Lyon,\\
Univ Lyon 1, CNRS,\\
Laboratoire de Physique,\\
F-69342 Lyon, France\\
\And
Pierre~Borgnat\\
Univ Lyon, Ens de Lyon,\\
Univ Lyon 1, CNRS,\\
Laboratoire de Physique,\\
F-69342 Lyon, France\\
\And
Patrice~Abry\\
Univ Lyon, Ens de Lyon,\\
Univ Lyon 1, CNRS,\\
Laboratoire de Physique,\\
F-69342 Lyon, France\\
\And
Ashish Bhaskar\\
Queensland University of Technology,\\
Smart Transport Research Centre,\\
Brisbane, Australia\\
\And
Edward~Chung\\
Hong Kong Polytechnic University,\\
Hong Kong
}

\maketitle

\begin{abstract}
Origin-Destination matrix estimation is a keystone for traffic representation and analysis. Traditionally estimated thanks to traffic counts, surveys and socio-economic models, recent technological advances permit to rethink the estimation problem. Road user identification technologies, such as connected GPS, Bluetooth or Wifi detectors bring additional information, that is, for a fraction of the users, the origin, the destination and to some extend the itinerary taken. In the present work, this additional information is used for the estimation of a more comprehensive traffic representation tool: the link-origin-destination matrix. Such three-dimensional matrices extend the concept of traditional origin-destination matrices by also giving information on the traffic assignment. Their estimation is solved as an inverse problem whose objective function represents a trade-off between important properties the traffic has to satisfy. This article presents the theory and how to implement such method on real dataset. With the case study of Brisbane City where over 600 hundreds Bluetooth detectors have been installed it also illustrates the opportunities such link-origin-destination matrices create for traffic analysis.
\end{abstract}


\section{Introduction}
\label{sec:Intro}

Massive urbanisation, as experienced nowadays, is placing urban infrastructure under ever increasing pressure. Whereas it was possible, in the seventies, for cities to satisfy growing demand by building more and more infrastructure, such solutions may no longer be possible nor appropriate -- space is becoming scarce and costs are rising. Hence, the optimisation of existing urban infrastructure has become an important challenge, which must be met with a good understanding of demand.

As such, traffic demand estimation has received much attention over the years, beginning in the sixties \cite{reilly1929methods}, and with extensive work undertaken since then. The aim of many studies is the estimation of the origin-destination (OD) matrices -- two entries tables that quantify zone-to-zone traffic demand. The estimation of such a matrix requires data for calibrating the estimation procedures. Traditionally, surveys have been used to sample the matrices directly. However, surveys are expensive, time consuming, biased, and their validity is limited in time \cite{veitch_whats_2013}. Consequently, methods based on data collected directly on roads have also been developed. Using such data, however, causes the OD matrix estimation problem to become a two-level problem: first, the estimation of the OD matrix and, second, its assignment to the network. The assignment consists in computing traffic flows on the road network that are consistent with the OD matrix and, thereby, enabling a comparison with collected data. 

Traffic counts are widely used as a data source for OD matrix estimation. Since the seventies, many cities have installed magnetic loops in road pavements that can detect the passage of massive metallic objects and convert the recorded electro-magnetic spikes into numbers of vehicles.

Considering the road network as a graph $\mathcal{G} = (V,L)$, where the set of $N_V$ vertices $V$ consists of road intersections (possible origin or destination) and the set of $N_L$ directed edges $L$ is the set of direct paths between intersections in $V$, the corresponding OD matrix is $\ddvar{T}$ of size $ (N_V)^2$. Magnetic loops, on links $l\in L$, produce $ N_L$ measures represented by the vector $\tilde{\q}$.
The generic problem of OD matrix ($\ddvar{T}$) estimation can be described as the following an inverse problem:
\begin{align}
\label{eqn:Intro:OD_InvProb}
(\widehat{\T},\ \widehat{\q}) & \in  \Argmin{\T,\,\q} \left\lbrace \gamma_1 \D_1(\tilde{\T},\T) + \gamma_2 \D_2(\tilde{\q},\q) \right\rbrace\\
 & \mbox{s.t.} \quad \q = F(\T)
\end{align}
where $\D_1$, $\D_2$ are two distance functions, and $\gamma_1$, $\gamma_2$ are two weights representing the relative belief in a prior knowledge of the OD matrix, $\tilde{\T}$ and the observed traffic counts $\tilde{\q}$, respectively. The assignment function, $F$, relates OD flows to road network links for comparisons against traffic counts.

Such a general formulation of the problem readily explains the extensive work in this area due to the many choices one can make for $\D_1$, $\D_2$, $F$ and for the \textit{a priori} OD matrix $\tilde{\T}$.

\subsection{Framework for OD matrices Estimation}

Faced with such an impressive and multi-faceted body of work, the point here is not to do a comprehensive literature review of all existing contributions. It is interesting, however, to note how traditional approaches have been formulated as Problem~\eqref{eqn:Intro:OD_InvProb}. Examples are gathered in Table~\ref{tbl:litrev} and in \cite{antoniou2016towards}.
\begin{table}[h]
\centering
\caption{Example of Distances used in Problem~\eqref{eqn:Intro:OD_InvProb}}
\label{tbl:litrev}
\begin{tabular}{lcl}
\toprule
Method Name & Distance Measure & References\\\midrule
Entropy maximisation & $\D_1(\T,\tilde{\T}) = \sum_{ij} T_{ij} ( \log T_{ij} - 1)$ & \cite{fisk1988combining,wilson2010entropy}\\
Information minimisation & $\D_1(\T,\tilde{\T}) = -\log \frac{ \prod_{ij} \left( \frac{{\tilde{T}}_{ij}}{\sum_{ij} {\tilde{T}}_{ij}}\right)^{T_{ij}}}{\prod_{ij} T_{ij}!}$ & \cite{van1980most,lam1991estimation}\\
Maximum Likelihood & $\D_1(\T,\tilde{\T}) =\sum_{ij} \left(\eta_{ij} T_{ij} - {\tilde{T}}_{ij}\ \log T_{ij} \right)$ & \cite{spiess1987maximum, ben1987methods}\\
Least Squares & $k\in[1,2],\quad \D_k(\tilde{x},x)=\frac{1}{2} \Vert \tilde{x}-x\Vert ^2$ & \cite{castillo2013bayesian,perrakis2015bayesian}\\
\bottomrule
\end{tabular}
\end{table}

Concerning the assignment, $F$, early approaches relied on the \textit{proportional assignment} assumption, that is, that demand ($\T$) is proportional to the link flows ($\q$). This is used in \cite{maher1983inferences,parry2012estimation}, amongst others. This assignment has the strong advantage of being simple; however, it is obviously inadequate when dealing with congested situations. 
In \cite{smock1962iterative} an iterative algorithm based on a cost-flow relationship, and aiming for Wardrop equilibrium, is proposed \cite{willumsen1978estimation}. Iterative solution for the computation of the assignment matrix is presented in \cite{yousefikia2016iterative}.

In \cite{yang1992estimation}, \textit{user equilibrium} is proposed, similar to that in \cite{cascetta1984estimation}, with a bi-level program where the \textit{lower level problem} is a deterministic user equilibrium assignment, and the \textit{upper level problem} is the estimation of the trip table (for example, based on the Generalised Least Squares approach). 
For more details on User Equilibrium approaches, the interested reader can refer to \cite{xu2018modified}.
In \cite{sherali1994linear}, a linear programming model is introduced. It estimates the flow for different paths, instead of the usual \textit{link} approach, in order to solve the deterministic user equilibrium assignment.
Variants of the deterministic user equilibrium assignment involve stochastic user equilibrium, as in \cite{yang2001simultaneous,maher2001bi}.
In \cite{codina2004adjustment}, the elastic demand traffic assignment is solved, using the notion of subgradient for cases where $\gamma_1 \D_1(\T,\tilde{\T}) + \gamma_2 \D_2(\tilde{\q},\q)$ would not be differentiable.

To directly account for the effects of congestion, \textit{Combined Distribution and Assignment} (CDA) models were developed with the aim of estimating the trip tables through a single objective function in which congestion is considered. In contrast to user equilibrium, where the assignment is implicitly defined (going from OD flows to link flows without intermediate trajectories), CDA, as in \cite{erlander1979calibration}, uses an assignment function.
The authors demonstrate, for congested networks, that the network equilibrium approach and the CDA approach give very similar results when the traffic counts correspond to a user equilibrium pattern.

Many studies have transposed these methods for time-dependent estimation \cite{toledo2015network}. Interested readers are referred to \cite{barcelo2015robust,antoniou2016towards, bauer2017quasi}.
With the new technologies, making vehicle identification possible, many of these methods have been revisited using such new sources of information for the computation of the prior OD matrix \cite{ge2016updating} and/or as calibration for the assignment model \cite{yang2017origin}. Opportunities offered by the new technologies also raised new concerns, as the finding of optimal sensor locations \cite{ye2016optimal}.

\subsection{Goal and contributions}
In the continuity of combined approaches, we propose in this paper to use the new traffic information to directly estimate an assigned OD matrix, that is, the Link-Origin-Destination (LOD) Matrix as introduced in \cite{michau2016primal}. The new technologies, as  Bluetooth detectors \cite{bhaskar2013fundamental}, offers the opportunity for trajectory retrieval \cite{michau2017bluetooth}, and to sample the LOD matrix. The real LOD matrix is then estimated by means of an inverse problem of dimension $N_V\times N_V \times N_L$, relying on traffic counts $\tilde{\q}$, and on a set of recovered trajectories. The inherent dimensionality of the problem remains identical to that of traditional approaches: despite the fact that  $\ddvar{T}$ is of size $(N_V)^2$, solving \eqref{eqn:Intro:OD_InvProb} is, in fact, an inverse problem of size $(N_V)^2 \times N_L$ because of the required assignment step, $F$, that actually routes each OD on links of the network.

Trajectory collection is now made possible by several new technologies such as GPS \cite{herrera2010evaluation}, Bluetooth \cite{michau2017bluetooth, hainen2011estimating, feng2015vehicle}, and Floating car data \cite{gomez2015evaluation}, among others.

\begin{figure}
		\begin{center}
		\subfloat[]{\includegraphics[width=7cm]{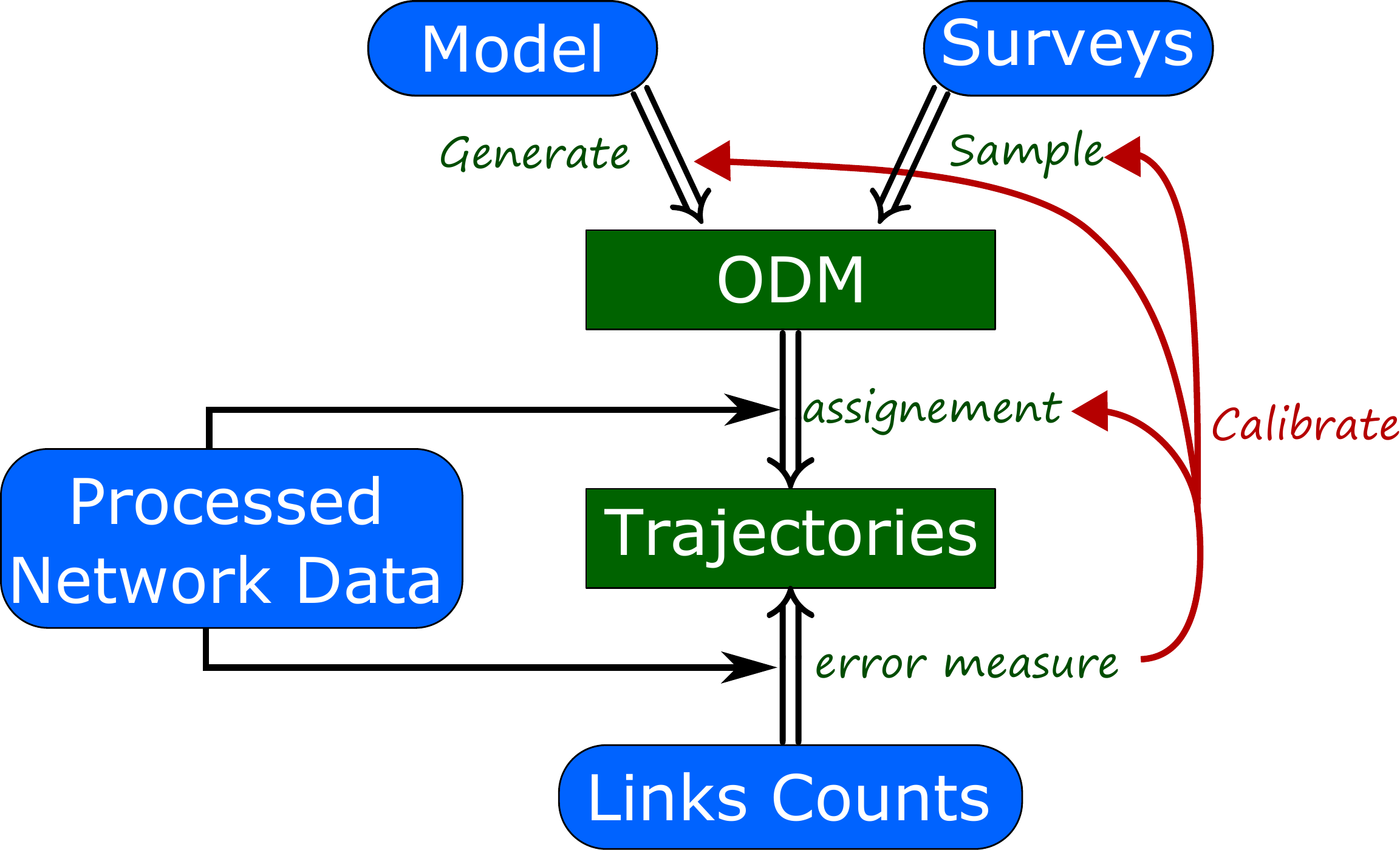}\label{sfig:TradFramework}}
		\subfloat[]{\includegraphics[width=7cm]{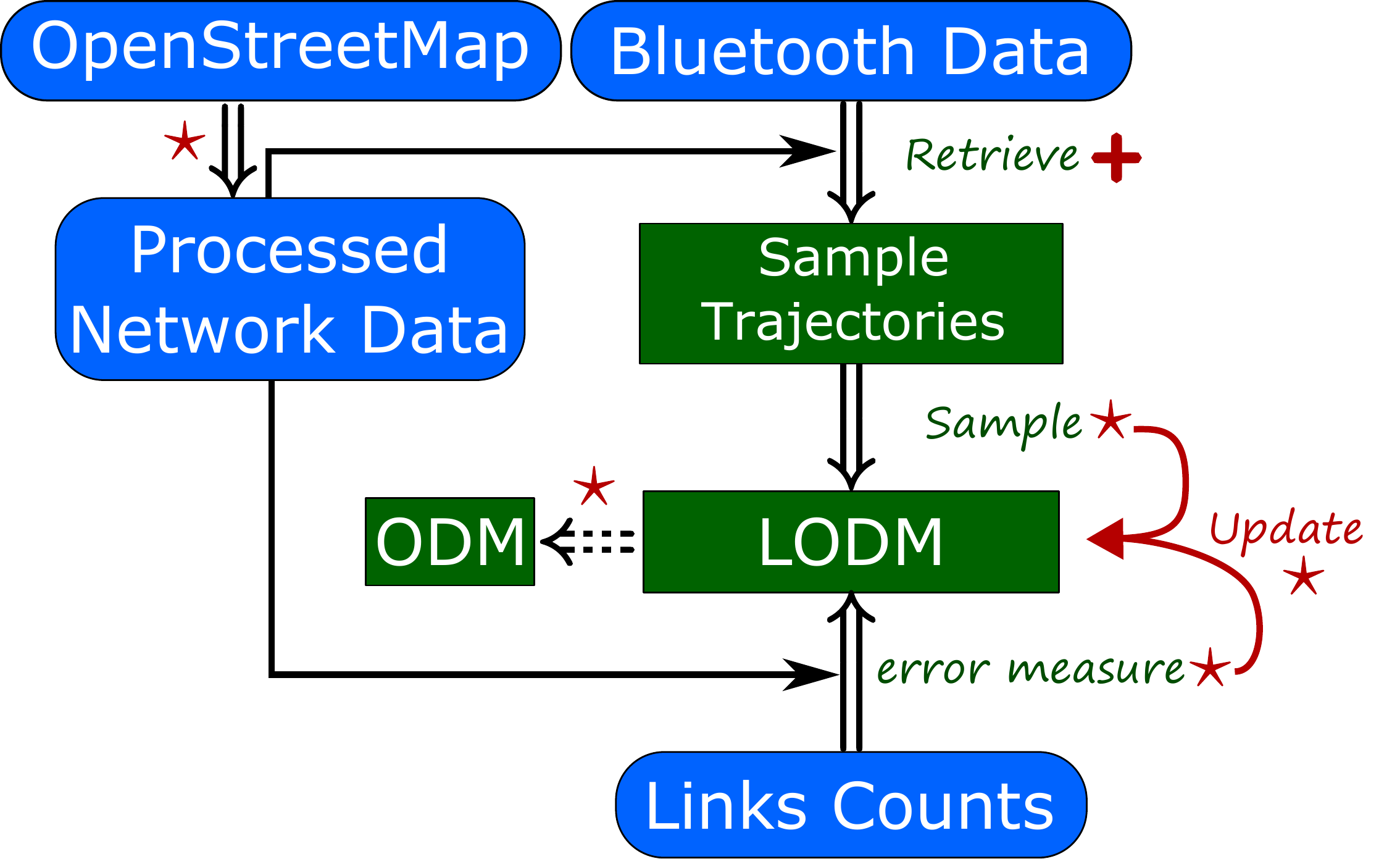}\label{sfig:ProposedFramework}}
		\caption{Traffic estimation: Traditional framework (a) versus proposed one (b). In blue boxes are inputs; in green boxes are traffic representation tools; in green font are models; and, in red are the feedbacks after comparison between model results and inputs. Red stars (\textcolor{red}{$\star$}) in (b) are processes detailed in this contribution. Trajectory retrieval (\textcolor{red}{\textbf{$\mathbf{+}$}}) is detailed in \cite{michau2017bluetooth}. The classical OD matrix is a by-product of the proposed method.}
		\label{fig:Methodology}
		\end{center}
	\end{figure}

The aim of this paper is to provide a clear and systematic framework for estimating the LOD matrix. It is based on a real case study, the city of Brisbane, because of the availability of both trajectories from Bluetooth data and of traffic counts from inductive loops. The methodology is illustrated in Figure~\ref{fig:Methodology}.

This paper is organised as follows: 
Section~\ref{sec:network} describes how the graph representing the road network, needed in all retrieval and estimation procedures, is built from raw data. It is based on our experience with the Brisbane case study. 
Section~\ref{sec:trajectories} provides a brief overview of how trajectories can be retrieved from Bluetooth data. 
Section~\ref{sec:lodm} illustrates how to implement the LOD matrix estimation procedure developed in \cite{michau2016primal} on this real dataset.
Finally, Section~\ref{sec:case_study} demonstrates the inherent potential of LOD matrices, proposing some applications for the results.

\noindent \textbf{Notation.} \quad 
	The following notation are used throughout this article: 
	$\udvar{X}$, $\ddvar{X}$ and $\tdvar{X}$ refer to vectors, matrices and tensors, respectively.
The Hadamard product (element-wise) of $\tdvar{Y}$ and $\tdvar{X}$ is denoted $\tdvar{Y}\circ \tdvar{X}$.
Subscript indices are used for dimensions over the nodes of the graph, and the index $i$ is used to label origins, $j$ to label destinations, and $k, m, n$ and $p$ to label nodes in general. Superscript indices are used for dimensions over the links and the indices $l$ and $e$ are favoured.

The symbol $\bullet$ is used to denote dimensions that do not contribute to a sum: \eg the sum over first and third dimensions, indexed with $i$ and $l$, is written $\sum_{i\bullet l} \tdvar{X}$.

We denote by $\Vert \cdot \Vert$ the element-wise norm for matrices: \eg $\Vert \ddvar{X} \Vert_1 = \sum_{ij} \vert X_{ij} \vert$ and $\Vert \ddvar{X} \Vert_2 = \left( \sum_{ij} X_{ij}^2 \right)^\frac{1}{2}$.

\section{Network Processing}
\label{sec:network}

	  	\subsection{Network Filtering}
	  	\label{Ssec:netw_filt}

Geo-referenced data (\eg traffic counts) need to be matched with road network concepts (\eg the concept of trajectory or shortest path). As such, the road network is an additional layer to the problem. 
To do so, it is customary to choose the GIS (Geographic Information System) representation, in which each road is represented by a geometrical object (most commonly a polyline, a connected sequence of straight line segments). This representation is convenient as the road network can then easily be interpreted as a graph $\G = (V,L)$.

	With a graph representation, traffic counts measure the number of vehicles using a road, that is, volumes on corresponding links (in $L$), and Bluetooth data, from Bluetooth Media Address Control Scanners (BMS) installed close to major intersections, and detecting vehicles without information as to their direction of travel or exact positions, can be interpreted as volume samples at the corresponding nodes.
	
	In this work, we use the road network from OpenStreetMap (OSM)\footnote{\url{https://www.openstreetmap.org/\#map=15/-27.4728/153.0268}}. It is freely available and has a good reputation for reliability. 
This layer, encompassing the Brisbane area (600km$^2$), is composed of around $62\,000$ roads, described with $432\,000$ coordinates (longitude, latitude) or $370\,000$ road segments. 

	GIS attributes are used here for filtering out non-essential roads (residential street, foot-way, cycleway, bus-way...). In this regard, it should be noted that traffic studies are mostly interested in traffic conditions on major roads, and consequently, it is where traffic data are collected. In the Brisbane case study, none of these minor roads are equipped with traffic detection devices. For the remaining road segments, a list of unique coordinates is extracted, denoted $V$, interpreted as the set of nodes of the graph. Then, the set of road segments, denoted $L$, characterises each road segment with the indices of its origin and destination nodes and its length. Note that the resulting graph is directed (as opposed to the OSM GIS representation, where the direction of travel is an attribute). As a result, the final number of links may be greater than the initial number of road segments. Some attribute information is retained at the node level: for instance, a flag indicates if the node belongs to a road in a tunnel or on a bridge.

	  	\subsection{Network Simplification}
	  	\label{Ssec:netw_simpl}
	
	From a traffic perspective, a fine description of the road geometry in between two intersections is unnecessary. Only the length is an important information. Removing nodes that are not intersections and that are not monitored by BMS, will simplify most of the graph analysis, reduce the problem size (as the LOD matrix has a size proportional to the number of links) without changing the traffic information retrieved by the methodology.

To remove such nodes, the following ad-hoc algorithm is proposed:
	Let us denote by $\G^\star = (V^\star,L^\star)$ the original road graph. 
Each element in $V^\star$ is of size $N_V \times 2$ (each node two coordinates) and each element in $L^\star$ is of size $N_L\times 3$ (for each edge: the starting and ending node index, and its length). 
In addition, let us define $S$, the set of BMS, (of size $N_S \times 2$ for BMS coordinates) and $r$, of size $N_S$ with each BMS detection range $r_s$ (the detection range could depend, for example, on whether the BMS is in a tunnel).
Finally, $\{\M^{V^\star}_r\}_{s\in S}$ is the mapping from the space of BMS $S$ to the space of nodes $V$ with parameters $r$, such that, $\M^{V^\star}_r(s)$ is the set of vertices in $V^\star$ within $r_s$ of the scanner $s\in S$.  
		
The simplified graph $\G = (V,L)$ is initialised as a copy of $\G^\star = (V^\star,L^\star)$. 
Let us also define $L_{Map}$ of size $\vert L^\star\vert$, a variable for keeping track of the link modifications, and initialised with zero values. 
$L_{Map}$ is given a `$-1$' value when an edge no longer appears in the resulting graph. If an edge has been combined with another link, its value in $L_{Map}$ is set to the index in $L^\star$ of this other link.
	
While any value has changed in $L_{Map}$, in the previous iteration, do the following:
	\begin{enumerate}[leftmargin=0.5cm]
		\item For every node $v\in V/\{\M^V_r\}_{s\in S}$:
			\begin{enumerate}
				\item If the node corresponds to a dead-end, (connected to only one link, or two links with a single other node), then those links are flagged for removal (\ie flag those links with `$-1$' in $L_{Map}$).
				\item If the node lies along a one-way road, that is, it has one entering link and one exiting link, then the incident link becomes the concatenation of both links, and the exiting link is flagged for removal (\ie flag the exiting link with the index of the incident link in $L_{Map}$).
				\item If the node is along a two-way road, that is, there are two instances of a pair of links shared with another node, then both incident links are concatenated with the corresponding exiting links, and the exiting links are, themselves, flagged for removal (\ie flag with corresponding index in $L_{Map}$).
				\item If the node is along a one-way road for one direction of travel, and is a dead-end for the other direction, it is then treated as in case (a) and (b).
			\end{enumerate}
		\item For every link $l\in L$:
			\begin{enumerate}
				\item If its origin is identical to its destination, then it is flagged for removal (`$-1$' flag).
				\item If  it has the same origin and destination as another link, the shorter is kept, and the other is flagged for removal (\ie flagged with the index of the retained link).
			\end{enumerate}
	\end{enumerate}
Once the algorithm achieves convergence, obsolete links are removed from the list $L$: $L\leftarrow L(L_{Map} = 0)$, and accordingly, nodes that do not appear in $L$ are removed from $V$. Finally, $L$ and $V$ are re-indexed to reflect the new cardinality.

	Note that this process ignores nodes in $\{\M^{V^\star}_r\}_{s\in S}$. This is justified as it keeps unchanged the road infrastructure around the Bluetooth detectors, where traffic is observed. 
Meanwhile, the array $L_{Map}$ enables us to keep track of modifications to the graph and, therefore, to the actual infrastructure. Hence, this process allows us, if needed, to adapt other GIS information to this new simplified graph.

\section{Retrieving Trajectories}
\label{sec:trajectories}

A method for retrieving trajectories from BMS data in real and simulated case studies has been presented in \cite{michau2017bluetooth}. The proposed framework consists in cleansing data from duplicated MAC (same MAC for different vehicles driving at the same time on the network) or twin MAC (several MAC for one vehicle). It then divides the dataset into sequences of detections representing unique trips. A first data analysis, as presented in \cite{michau2014retrieving}, can help to do some hypothesis on the mode of transport of the user. Then, for each remaining traffic relevant sequence, the algorithm iterates over every node within the scanning range of each detector, computing the only shortest path that could have created the observed detection pattern, taking possible overlapping detection area into account. This algorithm achieves an accuracy of $84\%$ in the real case study. 

The output from the algorithm is a list of trips, each composed of the links in $L$ used by a Bluetooth-equipped vehicle. These trajectories can be directly interpreted as a Bluetooth LOD matrix $\B$ of size $N_S\times N_S \times N_L $, where each trajectory adds $+1$ to the elements $B_{ij}^l$ for each link $l$ used in the trajectory, where $i$ and $j$ are, respectively, the first and last detector of the corresponding detection sequence.

The OD information obtained from $\B$ is valuable in its own right. In \cite{michau2014retrieving}, it has been shown that the 20 most important ODs (aggregated at neighbourhood level) were similar to those computed by an advanced transport model, the Brisbane Transport Strategic Model. This aggreagation is common in transportation analysis and mitigates the overlapping detection range problem as Bluetooth scanner with overlapping detection range are likely to belong to the same zone. In addition, multiplying $\B$ by an average penetration factor would lead to the solution usually proposed in the literature on ODM estimation using Bluetooth data \cite{carpenter2012generating}. It has been shown in \cite{michau2016primal}, however, that having a set of trajectories and link counts as inputs permits the direct estimation of Link-Origin-Destination Matrices (LODM) corresponding to the Origin-Destination Matrix already assigned to the network. Such results are achieved with a more advanced expansion of $\B$.

\section{Estimating the LOD Matrix}
\label{sec:lodm}

	\subsection{Existing Work and Differences}

The method proposed here demonstrates how to implement the theoretical work developed in \cite{michau2016primal} for real, road network datasets. 
In particular, the algorithm developed there requires modification because some assumptions underlying the theoretical work do not translate into the real world problem. 

First, traffic counts in a real network can, generally, only be monitored on a subset of roads, and not on every single road, as assumed in \cite{michau2016primal}. This is circumvented here by comparing observed and estimated counts on this subset only. 

Second, the assumption that every intersection is monitored by Bluetooth detectors might not be true. This depends, in particular, on the level of detail chosen for the network representation. It leads us to define the LOD matrix, similarly to the Bluetooth LOD matrix, as a matrix representing flows between Bluetooth detectors, that is a matrix in $S\times S \times L$. From another perspective, this could correspond to the addition of virtual links of weight $0$ from each Bluetooth scanner to every intersection within its detection range. This implies treating the BMS as virtual nodes in $V$. 
Thus, we will denote $\G =(V,L)$, the graph representing the road network extended with those virtual links, and with Bluetooth detectors. 
In particular, we have $S \subset V$.
Bluetooth detectors now become the origin and destination points considered in this study, but this could easily be adapted to any other collection of origin/destination patterns (\eg centroids or zones). 
This choice of origin/destination points is justified in \cite{wang2012understanding}, which defines detector-to-detector travel information as \textit{transcient} information, and demonstrates its usefulness.

The proposed method consists in solving a problem whose formulation is similar to that of Problem~\eqref{eqn:Intro:OD_InvProb}, adapted to the LOD matrix case. More specifically, the aim is to solve Eq.\eqref{eqn:LSODM:Obj}

\begin{equation}
\label{eqn:LSODM:Obj}
\resizebox{.9\hsize}{!}{$\widehat{\Q}  \in \Argmin{\Q} \left\lbrace \gamma_{TC} f_{TC}(\Q) + \gamma_{P} f_P(\Q) + \gamma_{C} f_C(\Q) + \gamma_{K}f_{K}(\Q) + \gamma_{TV} f_{TV}(\Q) \right\rbrace$}
\end{equation}
where $f_\cdot$ are convex functions, modelling properties the estimates should satisfy, some of them including the actual measured data (traffic counts in $f_{TC}$ and Bluetooth trajectories in $f_{P}$). They are detailed below. It is interesting for the reader to note that $f_P$ and $f_{TC}$ can be interpreted as an adaptation of the distance functions $D_1$ and $D_2$, in Problem~\eqref{eqn:Intro:OD_InvProb}, to the present case. The remaining functions constrain the set of solutions toward more consistency. $\gamma_\cdot$ are positive weights, applied to the objectives, to model their relative importance within the global objective.

	\subsection{Objective Function for the Real Case Study}
	\label{Ssec:CaseStudy:T2S:OF}

		\subsubsection{Notation and Definitions}
We have already defined the graph, representing the road network, as $\G =(V,L)$, where $V$ represents the $N_V$ nodes, and $L$ the $N_L$ node connections.
The network structure is represented through the incidence and the \textit{excidence} matrices, respectively, $\ddvar{I}$ and $\ddvar{E}$ of size $N_V \times N_L$. 
These matrices describe the relations between the nodes and the edges, such that, for every $(k,l)\in V \times L$:
\begin{equation}
	\begin{array}{ll}
	 I^l_k\ & = \left\lbrace 
		\begin{array}{ll}
		1 & \mbox{if the link $l$ is arriving to the node $k$}, \\
		0  & \mbox{otherwise},\\
		\end{array}
		\right. \\
		\\
	 E^l_k\ & = \left\lbrace 
		\begin{array}{ll}
		1 & \mbox{if the link $l$ is starting from the node $k$}, \\
		0  & \mbox{otherwise}.\\
		\end{array}
		\right. \\
	\end{array}
\end{equation}

Note that, in graph theory, it is customary to name the difference $(\ddvar{I} - \ddvar{E})$ as \textit{Incidence Matrix}; however, we need both matrices, separately, in this work.

Let us also define the tensors $\tdvar{I_1}$ and $\tdvar{I_2}$ (resp. $\tdvar{E_1}$ and $\tdvar{E_2}$), corresponding to the replication of $\ddvar{I}$ (resp. $\ddvar{E}$), such that,
\begin{equation}
	\begin{array}{rl}
		(\forall m \in V) \;  & (I_1)^l_{km} =  \left\lbrace
		\begin{array}{l l}
		1 & \mbox{if link $l$ is arriving to node $k$,}\\
		0 & \mbox{otherwise,}\\
		\end{array}
		\right.\\
		(\forall k \in V) \;  & (I_2)^l_{km} =  \left\lbrace
		\begin{array}{l l}
		1 & \mbox{if link $l$ is arriving to node $m$,}\\
		0 & \mbox{otherwise.} \\
		\end{array}
		\right.\\
	\end{array}
\end{equation}

$S$ is the set of $N_S$ BMS and $S \subset V$.
$\Q$ is the LOD matrix of size $N_S\times N_S\times N_L$. $\B$ is the Bluetooth LOD matrix of same size ($S\times S\times L$). 
We denote by $\q$ of size $N_L$ the variable representing the traffic volumes on each link, by $\widetilde{L}\subseteq L$ the subset of links on which traffic counts are measured, and by $\tilde{\q}$ the measured traffic counts.

Using this notation, we relate the LODM $\tdvar{Q}$ to the classical OD matrix $\ddvar{T}$, of size $N_S \times N_S$, where each element $T_{ij}$ contains the traffic flow from BMS $i$ to destination $j$, as follows:

\begin{equation}
	\label{eqn:ODM}
	\ddvar{T} = \sum_{\bullet\bullet l} \tdvar{E_1}\circ \tdvar{Q} = \sum_{\bullet \bullet l} \tdvar{I_2}\circ \tdvar{Q}.
\end{equation}
	
		\subsubsection{Traffic Count Data Fidelity $f_{TC}$}
		\label{Sssec:CaseStudy:T2S:OF:TC}
Note that if we define $\udvar{\delta_{\widetilde{L}}}$, the Kronecker delta vector of size $N_L$, such that 
\begin{equation}
	(\forall l\in L) \quad \left(\delta_{\widetilde{L}}\right)_l = 
	\begin{cases}
	1 \quad \mbox{if\ }l\in \widetilde{L},\\
	0 \quad \mbox{otherwise.}
	\end{cases}
\end{equation}
then, by definition, we have $\tilde{\q}$ satisfying:
\begin{equation}
\tilde{\q} = \udvar{\delta_{\widetilde{L}}} \circ \tilde{\q}
\end{equation}

The magnetic loops are usually subject to counting errors, which are modelled here by a noise $\udvar{\varepsilon}$. Hence, the measured quantity $\tilde{\q}$ reads:
\begin{equation}
	\label{eqn:TrafficAggr}
	\tilde{\q} = \udvar{\delta_{\widetilde{L}}} \circ (\q^\star + \udvar{\varepsilon})
\end{equation}
where $\q^\star$ contains the true traffic volumes. Moreover,
\begin{equation}
	\label{eqn:TrafficAggr-2}
	\q^\star = \sum_{ij\bullet} \Q^\star
\end{equation}
Therefore, Eq.~\eqref{eqn:TrafficAggr} becomes
\begin{equation}
	\label{eqn:TrafficAggr-3}
	\tilde{\q} = \udvar{\delta_{\widetilde{L}}} \circ \left( \sum_{ij\bullet} \Q^\star + \udvar{\varepsilon} \right)
\end{equation}
The noise $\udvar{\varepsilon}$ and the true traffic LODM ($\tdvar{Q}^\star$) are unknown. $\tdvar{Q}^\star$ is the quantity that is to be estimated and to do so, assuming that $\udvar{\varepsilon}$ is a random unbiased Gaussian noise, we look for the variable $\Q$ minimising the negative log-likelihood derived from Eq.~\eqref{eqn:TrafficAggr-3}, leading, in this case, to the traditional least square estimation \cite{bera2011estimation}. Thus, the first term of the global objective function to be minimised is the function $f_{TC}$:
\begin{equation}
	\label{eqn:Obj:TrafficCount}
	f_{TC}(\Q)   = \Vert \tilde{\q}  -  \udvar{\delta_{\widetilde{L}}} \circ \sum_{ij\bullet} \Q \Vert^2.
\end{equation}
Note that, this function provides also very good numeric performances with non Gaussian noise as demonstrated in~\cite{boulanger2018nonsmooth}.

		\subsubsection{Poisson Bluetooth Sampling Data Fidelity $f_P$}
		\label{Sssec:CaseStudy:T2S:OF:P}
$\B$ appears as a noisy version of $\Q^\star$ for which the noise level depends on the penetration rate of the Bluetooth Technology among road users. The relation between the tensors $\B$ and $\Q^\star$ can thus be modelled by a Poisson law, typically applied to counting processes, including in traffic anaylis \cite{ben1987methods, le2016brownian}. The negative log Poisson likelihood is generally a good choice as it allows for intensity dependent variance \cite{boulanger2018nonsmooth}.
\begin{equation}
	\label{eqn:Poisson}
	(\forall i,j,l \in V \times V \times L) \quad  B_{ij}^l = \mathcal{P} ((\eta_o)_{ij} Q_{ij}^{\star l}).
\end{equation}
where $\mathcal{P}$ is a Poisson law with parameter $(\eta_o)_{ij} Q_{ij}^{\star l}$.

As $\eta_o$ is not directly measurable, we introduce an approximation of this penetration rate, denoted $\eta$, and calculated as:
\begin{equation}
\label{eqn:ov-eta}
\eta = \frac{\sum \left( \udvar{\delta_{\widetilde{L}}} \circ \sum_{ij\bullet} \B \right)}{\sum \left( \udvar{\delta_{\widetilde{L}}} \circ \tilde{\q} \right)}
\end{equation}

The resulting objective, denoted $f_P$, models the negative log-likelihood associated with the Poisson model \cite{combettes2007douglas}:
\begin{equation}
	\label{eqn:Obj:Poisson}
	f_P(\Q) = \sum_{ijl}\psi\left(B_{ij}^l,\eta Q_{ij}^l \right)
\end{equation}
where
\begin{equation}
\label{eqn:psi}
\psi(B_{ij}^l,\eta Q_{ij}^l ) = 
\left\{
\begin{array}{ll}
	B_{ij}^l \log  \eta Q_{ij}^l +  \eta Q_{ij}^l & \mbox{if\ } \eta Q_{ij}^l>0 \mbox{\ and\ } B_{ij}^l>0, \\
	\eta Q_{ij}^l 								  & \mbox{if\ } \eta Q_{ij}^l\geq 0 \mbox{\ and\ } B_{ij}^l=0, \\
	+\infty 									  & \mbox{otherwise}.
\end{array}
\right.
\end{equation}

		\subsubsection{Consistency Constraint $f_C$}
		\label{Sssec:CaseStudy:T2S:OF:D}
The data consistency term ensures that the total flow should be greater than the flow of Bluetooth-enabled vehicles. This is achieved by constraining $\Q$ to belong to the following convex set $C$:
\begin{equation}
	\label{eqn:ConvexSet}
	C = \big\{\Q  = \big({Q}_{ij}^l\big)_{(ijl)\in S\times S\times L} \in \R^{N_S \times N_S \times N_L}\;\vert\;  {Q}_{ij}^l\geq {B}_{ij}^l\big\}.
\end{equation}
The corresponding convex function is the indicator $\iota_C$:
\begin{equation}
	\label{eqn:Obj:domain}
	f_C(\Q) = \iota_C(\Q) = 
	\begin{cases}
	0 &\mbox{\ if\ } \;\;\Q\in C, \\
	+\infty &\mbox{\ otherwise.}
	\end{cases}
\end{equation}

Note that it is customary to use indicator functions taking values in $[0;1]$. The choice here of a function with values in $[0;+\infty]$ is supported by two arguments: First, it would not make sense were this constraint not satisfied; hence, the $\infty$ penalty. Second, this enables us to restrain our study to the case $\gamma_{C}$ in $[0,1]$ (any other value for $\gamma_C$ would not change the value of $\gamma_C f_C$).

		\subsubsection{Kirchhoff's Law $f_K$}
		\label{Sssec:CaseStudy:T2S:OF:Kirch}
The conservation of the number of vehicles at each intersection in the network is an additional property that our solution should respect. The conservation should ideally be satisfied independently for each origin-destination pair. This can be written as a Kirchhoff's law:
\begin{equation}
(\forall k\in V\backslash S,\quad \forall (i,j)\in S\times S) \quad 	\sum_l E^l_k Q^l_{ij}  = \sum_l  I^l_k Q_{ij}^l.
\end{equation}
To accommodate possible errors in measured data, we relax the equality and it results in a convex function to be minimised:
\begin{equation}
	\label{eqn:CaseStudy:OF:Kirchhoff}
	f_K(\Q) = \sum_{\substack{ijk \\ i,j \in S\times S \\ k\in V \backslash S}} \left( \sum_{l} \left( I^l_k - E^l_k \right) Q^l_{ij} \right)^2.
\end{equation}

Note that we exclude the BMS nodes from Kirchhoff's law due to their particular status -- traffic does not pass through them; these act as sources or sinks only.

		\subsubsection{Total Variation $f_{TV}$}
		\label{Sssec:CaseStudy:T2S:OF:TV}

Finally, it has been shown that minimising the total variation of traffic with same origins and similar destinations, or with same destinations and similar origins leads to better solutions and is relevant for traffic \cite{michau2016primal}.

To better define the term ``similar'' origins or destinations used above, we propose to compute the shortest paths between all pairs of detectors and to consider as ``similar'', BMS closer than $300$m. This threshold is chosen for two reasons; firstly, it needs to be above the detection radius of the BMS, and secondly, it should not be too high as to limit the number of pairs on which to compute the total variation.

Thus, total variation can be expressed as:
\begin{equation}
\label{eqn:CaseStudy:OF:TVge}
\begin{array}{rl}
f_{TV}(\Q ) & = \sum_{\substack{i,i'\in S\times S \\ i \neq i' \\ d_{i,i'}<300m}}\sum_{\substack{j,l \\ j\neq i,i'}}   \omega_{ii'} \vert Q_{ij}^l - Q_{i'j}^l\vert \\
            & + \sum_{\substack{j,j'\in S\times S \\ j \neq j' \\ d_{j,j'}<300m}}\sum_{\substack{i,l \\ i\neq j,j'}}   \omega_{jj'} \vert Q_{ij}^l - Q_{ij'}^l\vert \\
\end{array}
\end{equation}
where $d_{ij}$ is the length of the shortest path from detector $i$ to detector $j$, and $\omega_{ij}$ is a weight defined as follows:
\begin{equation}
(\forall (i,j) \in S\times S;\quad i\neq j) \quad \omega_{ij} = e^{-\frac{d_{ij}}{d_0}}.
\end{equation}
Here, $d_0$ is $300m$.

We define the matrix $\ddvar{J}$ of size $N_P\times N_S$, where $N_P$ is the number of detector pairs in $P$ with shortest path shorter than $300m$ as,
\begin{multline}
\forall (i,j,p)\in S\times S \times S^2;\ p=i\cdot N_S + j, \\
\begin{array}{l}
	J(p,i) = \left\lbrace 
	\begin{array}{ll}
	- \omega_{ij} & \mbox{if } d_{ij}\leq 300m \\
	0 & \mbox{otherwise,}\\
	\end{array}\right.\\
	J(p,j) = \left\lbrace 
	\begin{array}{ll}
	+ \omega_{ij} & \mbox{if } d_{ij}\leq 300m \\
	0 & \mbox{otherwise,}\\
	\end{array}\right.
\end{array}
\end{multline}
Then, the total variation can be expressed as
\begin{equation}
\label{eqn:CaseStudy:OF:TV}
f_{TV}(\Q ) = \sum_{l}   \Vert \ddvar{J}^\top   \ddvar{Q}^l \Vert_1 + \sum_{l}   \Vert \ddvar{J}^\top   (\ddvar{Q}^l)^\top \Vert_1,
\end{equation}
where $\ddvar{Q}^l$ models the $l$-th extracted matrix from $\Q$. Its dimension is thus $N_S \times N_S $.

	\subsection{Algorithm}
With the above definitions of the five terms $f_{TC}$, $f_P$, $f_C$, $f_K$ and $f_{TV}$, the optimisation problem \eqref{eqn:LSODM:Obj} can be solved by a Proximal-Dual Algorithm as in \cite{michau2016primal}.

	\subsection{Simulations}
	
The proposed methodology has been tested in simulated environment, once with a simpler problem in~ \cite{michau2016primal}, and once on a simulation of Brisbane city center with Aimsun as in \cite{michau2017bluetooth}. These two case studies have the advantage of providing a ground truth against which to compare the results.

The first simulated case study consists in creating random networks with 50 nodes and an average connectivity of 6. For varied number of users ($10^5$ and $10^4$), random origin-destination pairs are drawn. Last, for each OD pair, a Bluetooth penetration rate is also randomly assigned. This case study demonstrated that in average, the proposed methodology improves standards solutions, such as uniform expansion \cite{carpenter2012generating}, by $30\%$ on the root mean square error indicators (RMSE), while giving a better estimate of the total number of users on the network.
These results have been confirmed by the Brisbane city center simulation.

In both cases, the proposed framework brought encouraging results by estimating consistent values for traffic, both in terms of OD volumes and in term of assignment. Yet, these encouraging results were performed over quite restricted networks, and with non complex assignment function: experiments demonstrated that in small networks the shortest path is almost exclusively used, even with simulation software.

In the following, we demonstrate that the methodology is suitable for larger networks (city scale) and presents consistent and interesting results.

\section{The Brisbane Case Studies}
\label{sec:case_study}

Combining the three problems described above, that is, network processing, trajectory retrieval and LOD matrix estimation, it is now possible to estimate the LOD matrix of a real network. In the following, we present the results obtained when these processes were applied in Brisbane, and give a more comprehensive interpretation of the LOD matrix concept.

For this case study, we chose to investigate traffic on a generic weekday, Tuesday, 28th July, 2014, during the morning peak from 6\am\ to 9\am.

\subsection{Network}

As mentioned earlier, the Brisbane network, with which we will be working, is composed of around $62\,000$ roads, and described with $432\,000$ coordinates (longitude, latitude) and $370\,000$ road segments. 

After using attributes to filter out major roads, as discussed in Section~\ref{Ssec:netw_filt}, the Brisbane case study becomes a dataset with $14\,000$ roads ($96\,000$ undirected road segments) and  $110\,000$ coordinates. Then, interpreting as a graph, that is, with unique coordinates and directed links, we find that $|V|=N_V= 78\,000$ nodes, and $\vert L\vert=N_L = 121\,000$ links.

The simplification procedure of Section~\ref{Ssec:netw_simpl},  with parameter $r$ set to 150m, requires 13 iterations of the \texttt{`while'} loop, and yields a graph of size $N_V = 8\,900$ and $N_L = 18\,300$, with $2\,300$ nodes in $\{\M^V_r\}_{s\in S}$.
	
	\subsection{Size of the Problem}

The Brisbane metropolitan area had 576 Bluetooth detectors on 28th July 2014.
Restricting the study to the city centre only, or 225 km$^2$, the case study is composed of 430 Bluetooth detectors, 1\,800 intersections and 3\,950 links.
There are 3\,030 identified counting sites, composed of around 8\,000 magnetic loops, on the original road network.
During the simplification step presented in Section~\ref{Ssec:netw_simpl}, when links are concatenated, the resulting link is assigned the average of the counts (average for links with traffic count information only). 
When two links have same origin and destination, one is removed, and the remaining link is assigned the sum of the counts.
This process results in 1\,430 links with traffic count values (36\% of the links).

\begin{figure}
\hspace{-1cm}
		\subfloat[]{\includegraphics[height=5.8cm]{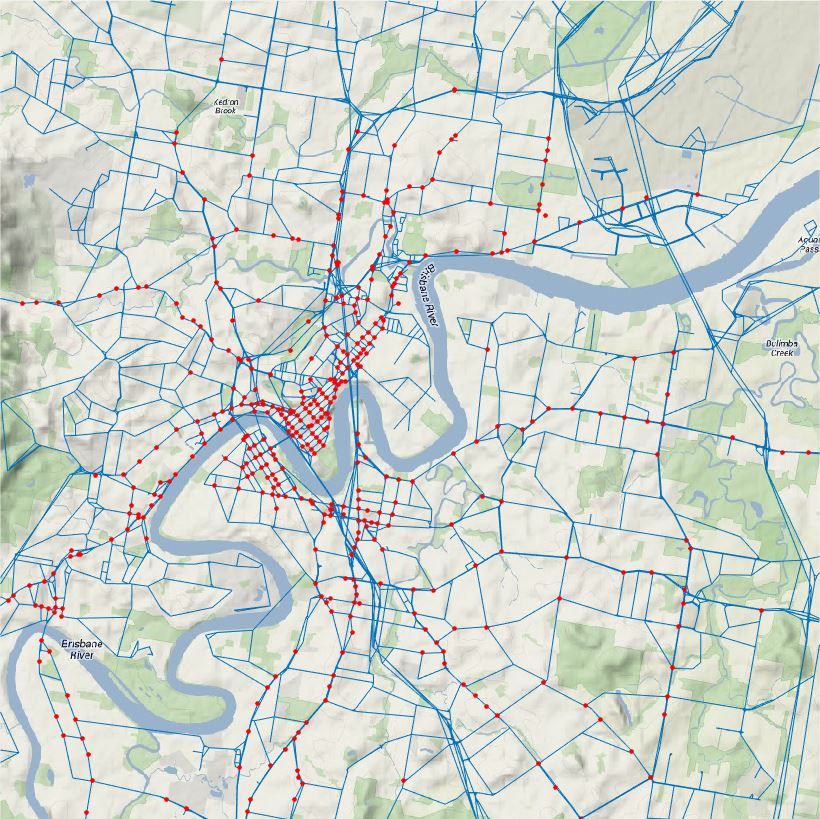}
						\label{sfig:CaseStudy_SimpNetw}}
		\subfloat[]{\includegraphics[height=5.8cm]{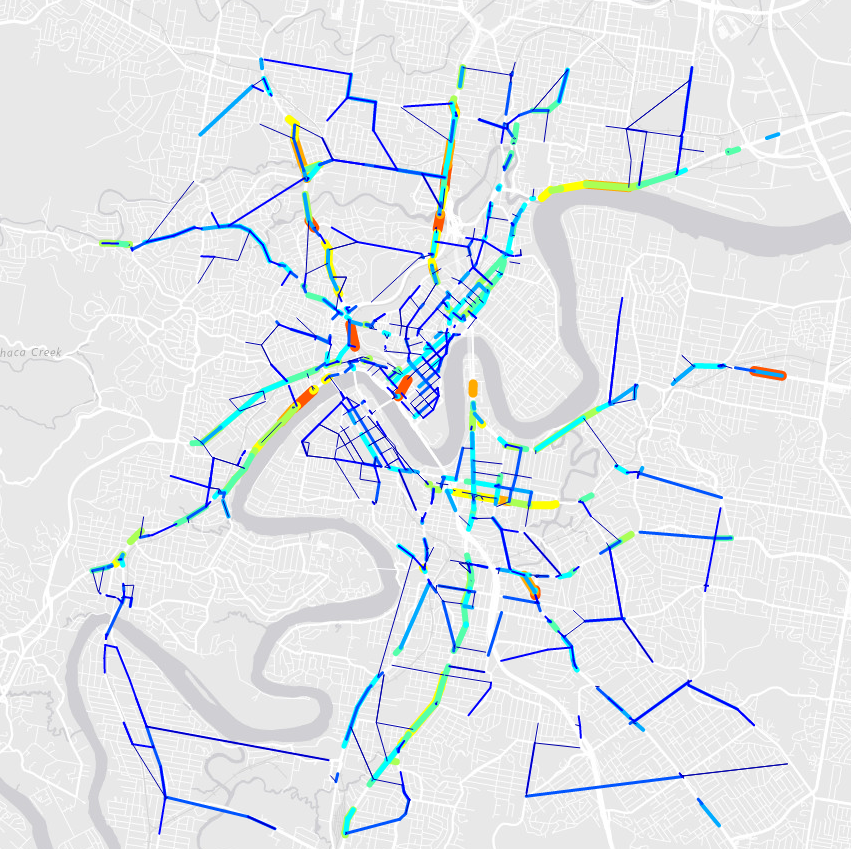}
						  \includegraphics[height=5.8cm]{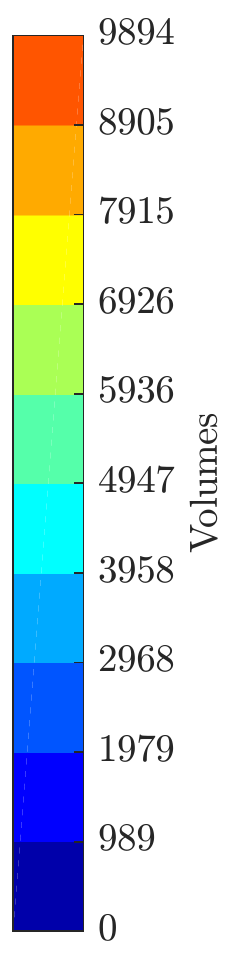}
					\label{sfig:CaseStudy_TrafficCounts}}
		\subfloat[]{\includegraphics[height=5.8cm]{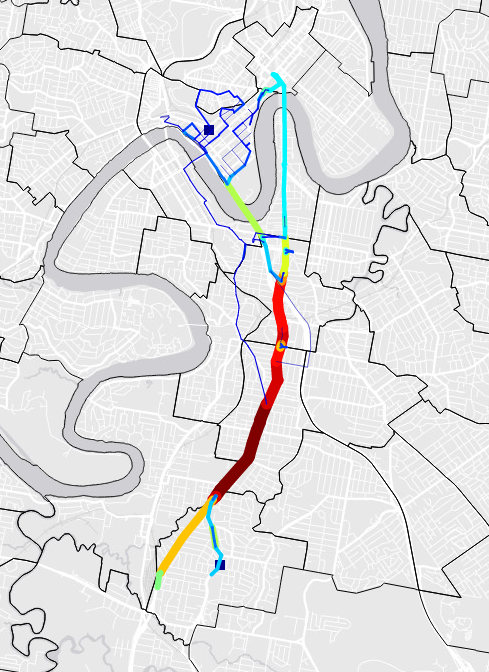}
						  \includegraphics[height=5.8cm]{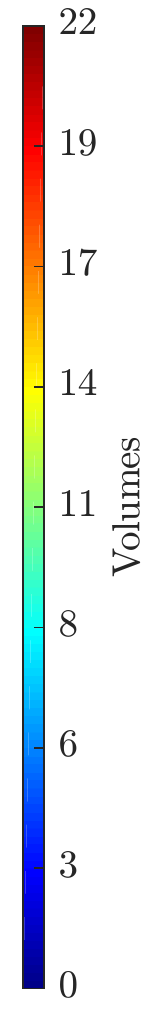}
				    \label{sfig:CaseStudy_1OD_Links_B}}
		\caption{(a) Traffic counts derived from induction loop detectors after transfer to the simplified network. Only 36\% of the links have non-zero values. (b) Bluetooth LOD flows for one specific OD pair (Brisbane CBD (upper \textcolor{blue}{$\blacksquare$}) to Moorooka (lower \textcolor{blue}{$\blacksquare$}). Colour and width of the roads represent traffic volume.}	
\end{figure}

Once virtual links and Bluetooth detectors are considered, the graph $\G=(V,L)$ is of size: 
$N_V=2\,230$, $N_L = 5\,370$, $N_S=430$, $N_{\widetilde{L}} = 1\,430$, and $N_P = 1\,190$ (pairs of Bluetooth detectors with paths shorter than 300m). The simplified network is illustrated in Figure~\ref{sfig:CaseStudy_SimpNetw}.
For the 6 \am\ to 9 \am\ time interval, the Bluetooth LOD matrix is composed of 39\,100 trajectories.
The cumulated number of traffic counts is 3\,252\,172.
The Bluetooth OD penetration rate, $\eta$, as per Equation~\eqref{eqn:ov-eta}, is 0.21.
The total number of vehicles is unknown. 

\figref{sfig:CaseStudy_TrafficCounts} illustrates the traffic count values for roads in $\widetilde{L}$ during the morning peak hours, and \figref{sfig:CaseStudy_1OD_Links_B} presents, for one OD (Brisbane CBD to Moorooka), the road traffic as recovered from the Bluetooth data.
Computations were done with a 2012 laptop equipped with a core i7 intel processor, 16 GB of RAM.The proximal primal dual algorithm is implemented with Matlab and it was left running for 5 days. The two last consecutive estimations of the LOD matrix have an RMS-Difference below $10^{-3}$.

	\subsection{Estimation and Assessment of Brisbane LODM}

For this case study, a complete knowledge of actual traffic flows is not available; therefore, assessment metrics comparing results to reality cannot be used. 
Yet, achieved solutions can be compared with naive solutions, using the values of each term of the objective function as indicators of the solution consistency.
Additionally, we are able to discuss the results of two different solutions of the inverse problem~\eqref{eqn:LSODM:Obj}, computed for the two best set of parameters $\lbrace\gamma_\cdot\rbrace$ identified in \cite{michau2016primal}.

In \tblref{tbl:CaseStudy:Solution}, we compare the indicators for these two solutions and for a naive solution, $\widehat{\Q}^0$, defined as:
\begin{equation}
\widehat{\Q}^0 = \frac{1}{\eta}\cdot \B
\end{equation}
Note that this solution can be interpreted as the equivalent, for an LOD matrix, of the solution usually found in the literature for OD matrix estimation from datasets with vehicle identification \cite{carpenter2012generating}.

For the three estimates, we also compare the total number of vehicles in the LOD matrix, computed with the benefit of the two relationships in Equation~\eqref{eqn:ODM}. Indeed, the total number can be obtained either, by counting vehicles leaving every origin or, by counting vehicles arriving at every destination. This leads to two estimations of the total number of vehicles, $N_{(Or)}$ and $N_{(Dest)}$. If Kirchhoff's law is perfectly satisfied, one should have $N_{(Or)} = N_{(Dest)}$.

\begin{table}
\centering
\caption[LOD Matrix Estimation: Results for Brisbane Case Study]{LOD Matrix Estimates Comparison. $\gamma_P = \gamma_C = 1$.}
\label{tbl:CaseStudy:Solution}
\begin{tabular}{lllllllllll}
\toprule
         			& $\gamma_{TC}$ & $\gamma_{K}$ 	& $\gamma_{TV}$ & & $f_{TC}$ & $f_K$ 	& $f_P$ 	& $f_{TV}$ 		& $N_{(Or)}$	& $N_{(Dest)}$	\\
\midrule
$\widehat{\Q}^0$	&				&				&				& & 162	 	 &  0		& 0.0025	& 19\,050\,400	& 186\,181		& 186\,181	\\
$\widehat{\Q}_1$		& 17.78			& 0.0128		& 0.0212		& & 54	   	 & 922		& 12201		& 13\,904\,421	& 162\,090	& 165\,654	\\
$\widehat{\Q}_2$		& 1.78			& 0.228			& 0.0377		& & 134	   	 & 53		& 76976		& 12\,906\,754 	& 156\,401	& 159\,911		\\
\bottomrule
\end{tabular}
\end{table}

\begin{figure}
	\centering
	\subfloat[]{\includegraphics[height=\hsize/2]{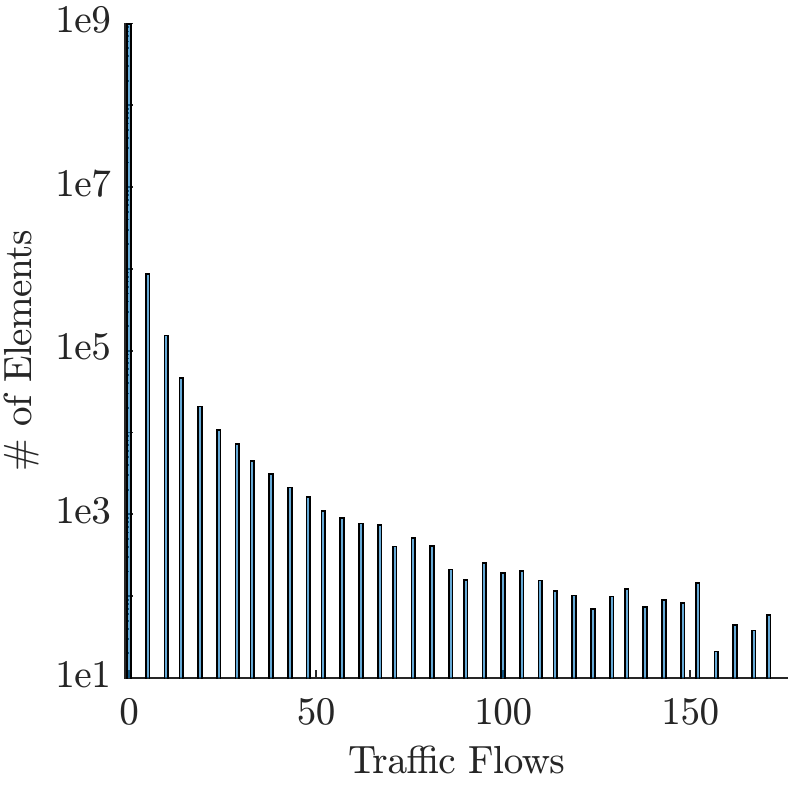} \label{sfig:CaseStudy_Histo_Q0}}
	\subfloat[]{\includegraphics[height=\hsize/2]{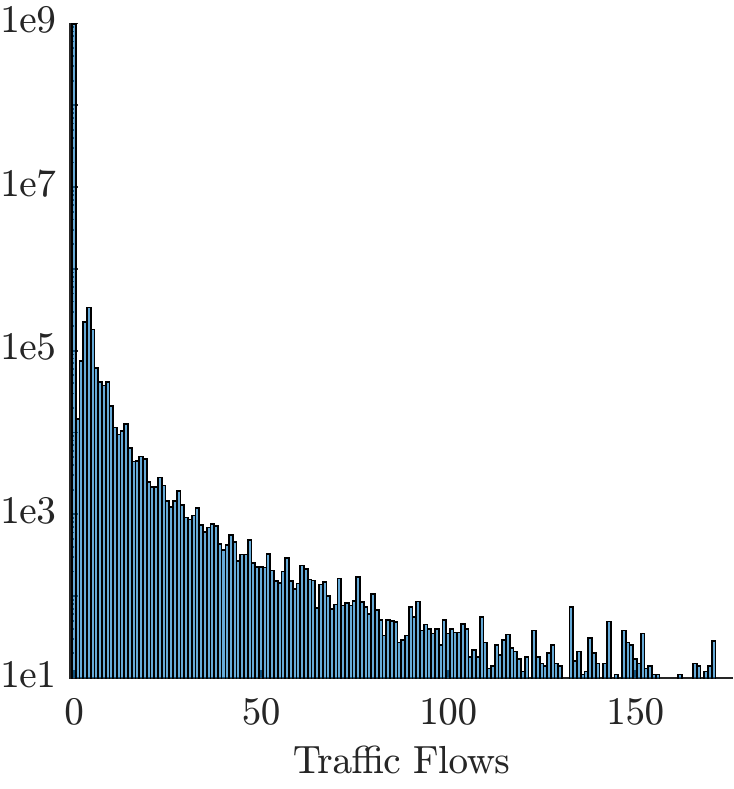} \label{sfig:CaseStudy_Histo_Q}}%
	\caption{Distribution of traffic flows in the LODM, (a) for $\widehat{\Q}^0$, (b) for the estimated LOD matrix $\widehat{\Q}_1$. The Y-axis is logarithmic. In (a), the forced quantization of the flow values is apparent.}
		\label{fig:CaseStudy_Histo}		
\end{figure}

The solution $\widehat{\Q}^0$ appears to performs well on the indicators $f_{TC}$, $f_K$ and $f_P$ when compared to the other solutions. This is to be expected as, by its design, this solution is based on measured Bluetooth trajectories (hence, the perfect satisfaction of Kirchhoff's law), multiplied by the average penetration factor (hence, the good performance on $f_P$). This global penetration factor is computed using the measured traffic counts, hence the good performances on $f_{TC}$.
Yet, we have shown in \cite{michau2016primal} that this solution is not satisfactory: it only estimates flows on roads where Bluetooth samples are available. Moreover, the estimated flows are, by design, restricted to multiples of the global penetration rate only (\cf \figref{sfig:CaseStudy_Histo_Q0}).
Because the traffic flows between BMS are usually small integer numbers (\eg Fig~\ref{sfig:CaseStudy_Histo_Q}), this limitation on the values taken by traffic flows makes the estimation of low value OD flows not efficient. As a consequence, the estimated number of vehicles on the road networks is higher than for other solutions.
\noindent

The other estimates, denoted by $\widehat{\Q}_1$ and $\widehat{\Q}_2$, represent trade-offs between the satisfaction of the four relaxed properties from which the objective function was built. Note that the consistency constraint $f_C$ is always satisfied. Each weight $\gamma_\cdot$ represents the importance accorded to the satisfaction of the associated property. For example $\widehat{\Q}_2$ has a relatively higher $\gamma_K$ value, and thus satisfies Kirchhoff's law better than $\widehat{\Q}_1$. Against this, $\widehat{\Q}_1$ performs better on $f_{TC}$ and $f_P$ values.

Given that a complete and accurate picture of the actual traffic flows is not available here, it is impossible to measure which of the solutions is best. 
Nevertheless, we believe that estimating traffic flows on the whole network, rather than on sampled trajectories only, provides a more interesting insight into traffic in the city. 
Moreover, at present, comparison with existing literature does not bring much additional insight. 
In any case, comparing the resulting OD matrices gives satisfying results, as does the direct comparison between the Bluetooth-measured OD matrix and the literature \cite{michau2014retrieving}. 
Comparing the estimated assignment to that of other works, usually derived from a model, would therefore provide no information as to how well it matches actual traffic on the day. These questions are thus left unanswered and will be the subject of future work.

Notwithstanding this inability to test the accuracy of the estimated LOD matrix, a qualitative study of the results can nonetheless help to bolster confidence and demonstrate some of the benefits it delivers. In the following, we will use the solution $\widehat{\Q}_1$ to showcase traffic information that can be inferred from the LOD matrix of Brisbane.

	\subsection{Example of a Traffic Map from the LOD Matrix}
\begin{figure}
	\centering
		\subfloat[]{\includegraphics[height=6cm]{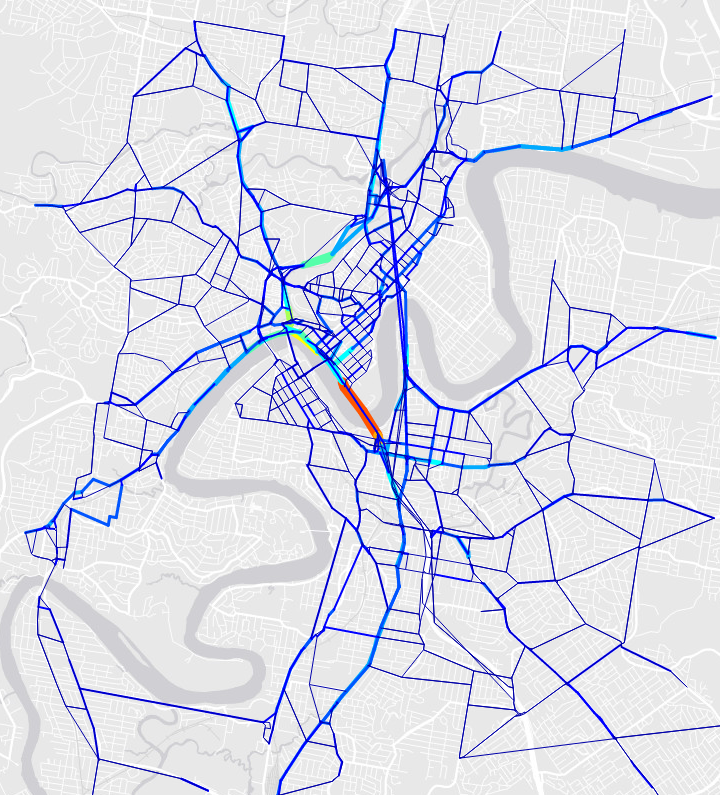}%
					\includegraphics[height=6cm]{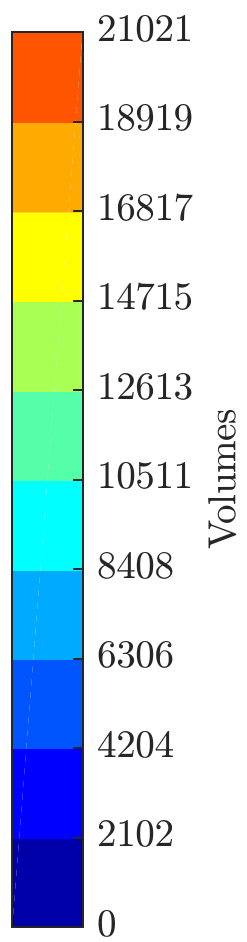}\label{fig:CaseStudy_LODM_TrafficCount}}
		\subfloat[]{\includegraphics[height=6cm]{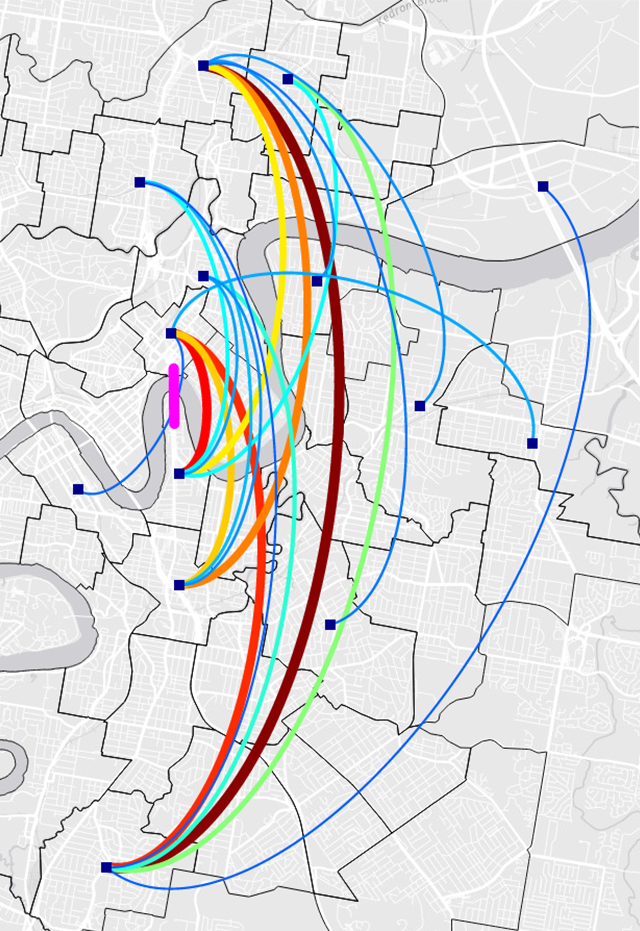}%
		 			\includegraphics[height=6cm]{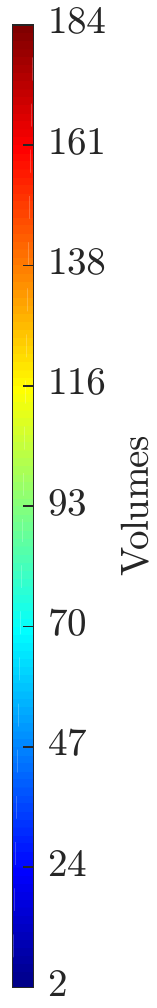}\label{fig:CaseStudy_1Link_ODs}}%
		\subfloat[]{\includegraphics[height=6cm]{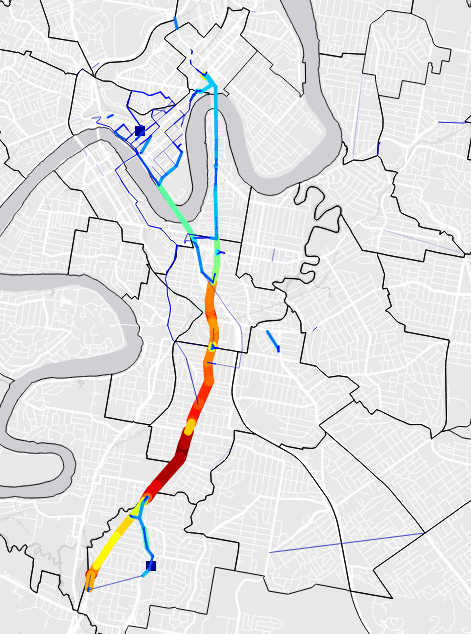}%
					\includegraphics[height=6cm]{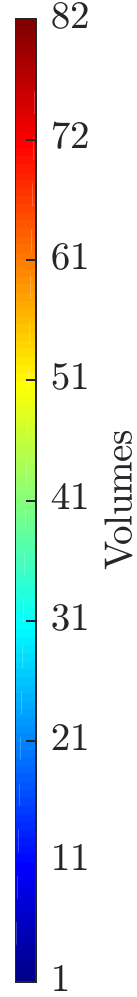}\label{fig:CaseStudy_1OD-links}}
		\caption{(a) Brisbane Traffic Counts on the simplified network for the area of study. (b) Origin-Destination flows of vehicles using the Bradfield Highway Bridge (North to South direction). The bridge is the road highlighted in magenta. Width and colour of the semi-ellipses are alternatively used to indicate OD volumes. Only the twenty largest OD volumes are shown. (c) Road volumes for vehicles trips from CBD (upper \textcolor{blue}{$\blacksquare$}) to Moorooka (lower \textcolor{blue}{$\blacksquare$}) in Brisbane. Some isolated links have non-zero traffic flows: These artefacts stem from the term $f_{TC}$ that favours non-zero flows on links where observed traffic counts are high. A stronger weight on the Kirchoff's law term in the objective function would remove such artefacts at the expense of other objectives.}
		\label{fig:10}		
\end{figure}

Firstly, link volumes for the whole road network can be mapped, similarly to \figref{fig:CaseStudy_LODM_TrafficCount}. Compared to the measured traffic counts shown in \figref{sfig:CaseStudy_TrafficCounts}, these flows are more continuous and consistent on adjacent roads.  


As a further illustration, one can choose a road segment and extract its corresponding OD matrix to obtain a better understanding of road usage. 
In \figref{fig:CaseStudy_1Link_ODs}, the Bradfield Highway Bridge is selected (in magenta), and the twenty most important OD flows are represented. For a more readable map, OD flows are aggregated by Statistical Local Areas (SLA).
This figure shows that a sizeable number of vehicles uses the Bradfield Highway Bridge to cross Brisbane; indeed, the OD flow from the northernmost SLA to the southernmost SLA is most the important.
Another identifiable behaviour is that an important fraction of vehicles in the eastern part of Brisbane use this bridge rather than the Gateway motorway bridge, at the far east of the river, even though this is not the fastest path. In fact, this bridge is the easternmost toll-free bridge, and drivers might prefer it to a tolled alternative.


Finally, approaching the data another way, it is possible to extract for any OD pair the corresponding road usage. In \figref{fig:CaseStudy_1OD-links}, two SLA regions are selected (Brisbane CBD to Moorooka), and traffic volumes on each link, for this OD only, are represented proportionally with colour, and alternatively, width. If this figure were plotted for $\widehat{\Q}^0$, it would be exactly similar to \figref{sfig:CaseStudy_1OD_Links_B} with all the flows multiplied by the global penetration rate $4.76$. In \figref{fig:CaseStudy_1OD-links}, however, the flows have been multiplied by different penetration factors. For example, the largest flows have been multiplied by less than 4, and the small traffic flows in South Bank (the SLA, west and across the river compared to the CBD) by a factor of 3 only. 
This illustrates the strong dependency of penetration rates on OD pairs and chosen paths.

\section{Conclusion}

This article described the major steps necessary for estimating the LOD matrix from the raw data inputs: a GIS road network, Bluetooth detections and traffic counts. After detailing how a GIS road network can be reduced to a relatively small graph (from $370\,000$ road segments to $18\,300$ links, or to $5\,370$ if restricted to the city centre), this article described how to implement a method for LOD matrix estimation in a real, large-scale case study. 
It demonstrated that the method is suitable for application to a large network ($430^2\simeq 185\,000$ OD pairs, $5\,370$ links), and in an urban context (20x20 km$^2$ in Brisbane City).

In addition, we provided some insights as to why the estimated solutions are more satisfactory than naive estimates. Yet, we were unable to analyse further the efficiency of the method in the absence of more comprehensive, actual traffic data. We believe that a more detailed investigation into the efficiency of this method, using real data, would make a worthwhile future contribution. In particular, the question of selecting the most adapted $\gamma_\cdot$ values remains an important, unanswered question, and is left for future study.

By illustrating several possible uses of the LOD matrix for traffic analysis, this article has proven that the concept works, and has highlighted its potential. 
Without resorting to any additional model or estimation procedure, OD flows and road usage can be jointly analysed and represented, so providing a more detailed understanding of traffic than the traditional OD matrix.
We have presented several useful ways of interpreting traffic information, which were made possible by the access to the LOD matrix; yet, the aim has not been to undertake a complete assessment of such applications, and many others could be developed, amongst others, traffic evolution by comparing successive estimations, commuter traffic analysis comparing morning and evening peak periods.

\printbibliography

\end{document}